\documentclass[twocolumn,twoside,10pt,unsortedaddress]{revtex4-1}
\usepackage{graphicx}
\usepackage{fancyhdr}
\usepackage{amsfonts}
\usepackage{amsmath}
\usepackage{amssymb}
\usepackage{xspace}
\usepackage{multirow}
\usepackage{floatflt}
\usepackage{tabularx}
\usepackage[pdftitle={G-Piano-Cyg_X-3}, colorlinks=true, citecolor=blue, linkcolor=blue, urlcolor=blue, pdfauthor={Giovanni Piano}]{hyperref}

\def \gray {$\gamma$-ray\xspace}
\def \grays {$\gamma$-rays\xspace}
\def \flx {photons $\mathrm{cm}^{-2}$ $\mathrm{s}^{-1}$}
\def \grid {AGILE-\textit{GRID}\xspace}
\def \lat {\textit{Fermi}-LAT\xspace}

\begin{document}

\title{AGILE monitoring of the microquasar Cygnus X-3}

%

\author{G.~Piano}\email[e-mail:~]{giovanni.piano@iasf-roma.inaf.it}
\author{M.~Tavani}
\author{V.~Vittorini}
\affiliation{INAF/IASF-Roma, Via del Fosso del Cavaliere 100, I-00133 Roma, Italy}
\author{A.~Bulgarelli}
\affiliation{INAF/IASF-Bologna, Via Gobetti 101, I-40129 Bologna, Italy}
\author{on behalf of the AGILE Team}
\affiliation{ }
\author{M.~McCollough}
\affiliation{Smithsonian Center for Astrophysics, 60 Garden Street, Cambridge, Massachusetts 02138, USA}
\author{G.~Pooley}
\affiliation{Astrophysics Group, Cavendish Laboratory, 19 J. J. Thomson Avenue, Cambridge CB3 0HE, UK}
\author{S.~Trushkin}
\affiliation{Special Astrophysical Observatory RAS, Karachaevo-Cherkassian Republic, Nizhnij Arkhyz 369169, Russia}

\begin{abstract}
AGILE data on Cygnus X-3 are reviewed focussing on the correlation between the production of \gray transient emission and spectral state changes of the source. AGILE clearly establishes a relation between enhanced \gray emission and the ``quenched'' radio/hard X-ray states that precede in general major radio flares. We briefly discuss the theoretical implications of our findings.
\end{abstract}

\maketitle

\thispagestyle{fancy}


\section{Introduction}
Cygnus X-3 is a microquasar discovered, as a bright X-ray source, in 1966 \citep{giacconi_67}. The companion star is a Wolf-Rayet star with a strong stellar helium wind ($\dot{M} \sim 10^{-5} M_{\odot}\mathrm{y}^{-1}$, $v_{wind} \sim 1000~\mathrm{km~s^{-1}}$). The system is located at a distance of about 7-10 kpc and the orbital period is 4.8 hours, inferred from infrared \citep{becklin_73}, X-rays \citep{parsignault_72} and \grays \citep{abdo_09}. Due to the very tight orbit ($d\approx3\cdot10^{11}$ cm), the compact object is totally enshrouded in the wind of the companion star. The nature of the compact object is still unknown \citep{vilhu_09}: published results suggest either a neutron star of 1.4 $M_{\odot}$ \citep{stark_03} or a black hole with a mass $\lesssim 10~M_{\odot}$ \citep{hanson_00}. In the radio band the system shows strong radio flares (``major flares'') reaching up to few tens of Jy. Radio observations at milliarcsec scales confirm emissions (at cm wavelengths) from both a core and a one-sided relativistic jet ($v \sim 0.3-0.7c$), with an inclination to the line-of-sight of $\lesssim14^{\circ}$ \citep{mioduszewski_01}. The radiation from the jet dominates the radio emission from the core during (and soon after) the major flares \citep{tudose_10}.

Cygnus X-3 exhibits a clear and repetitive pattern of (anti)correlations between radio and X-ray emission, and an overall anticorrelation between soft and hard X-ray fluxes \citep{mccollough_99,szostek_08}.

Gamma-ray detections of Cygnus X-3 were reported in the 1970s and in the 1980s at TeV \citep{danaher_81,lamb_82,vladimirsky_73} and PeV energies \citep{bhat_86,samorski_83}. However, subsequent observations by more sensitive ground-based telescopes did not confirm TeV and PeV emission from the source \citep{oflaherty_92}. Furthermore, the \textit{COS-B} satellite could not find any clear emission from Cygnus X-3 at MeV-GeV energies \citep{hermsen_87} and \textit{CGRO}/EGRET observations of the Cygnus region (1991-1994), even confirming a \gray detection above 100 MeV consistent with the position of Cygnus X-3 \citep{mori_97}, could not demonstrate a solid association with the microquasar. The firm \gray detection of Cygnus X-3 was announced at the end of 2009: on December 2, 2009 the AGILE Team claimed the discovery of strong \gray flares above 100 MeV \citep{tavani_09}, and on December 11, 2009 the \lat collaboration confirm the \grid results, by announcing the firm detection in \grays of the 4.8-hour orbital period of Cygnus X-3 \citep{abdo_09}.

Here we present an extended analysis on Cygnus X-3 with respect to the work published by the AGILE Team in 2009 \citep{tavani_09}, taking into account the \grid data collected between 2007-November-02 and 2009-July-29, during the AGILE ``pointing'' mode data-taking. We note the temporal repetitive coincidence of the \gray major flares with peculiar soft X-ray spectral states and pre-flaring radio states. We briefly discuss the theoretical implications of our findings in the perspective of the spectral modeling of this microquasar.

\section{Observations and data analysis}
\begin{figure}[h!]
 \begin{center}
	\includegraphics[width=8.0cm]{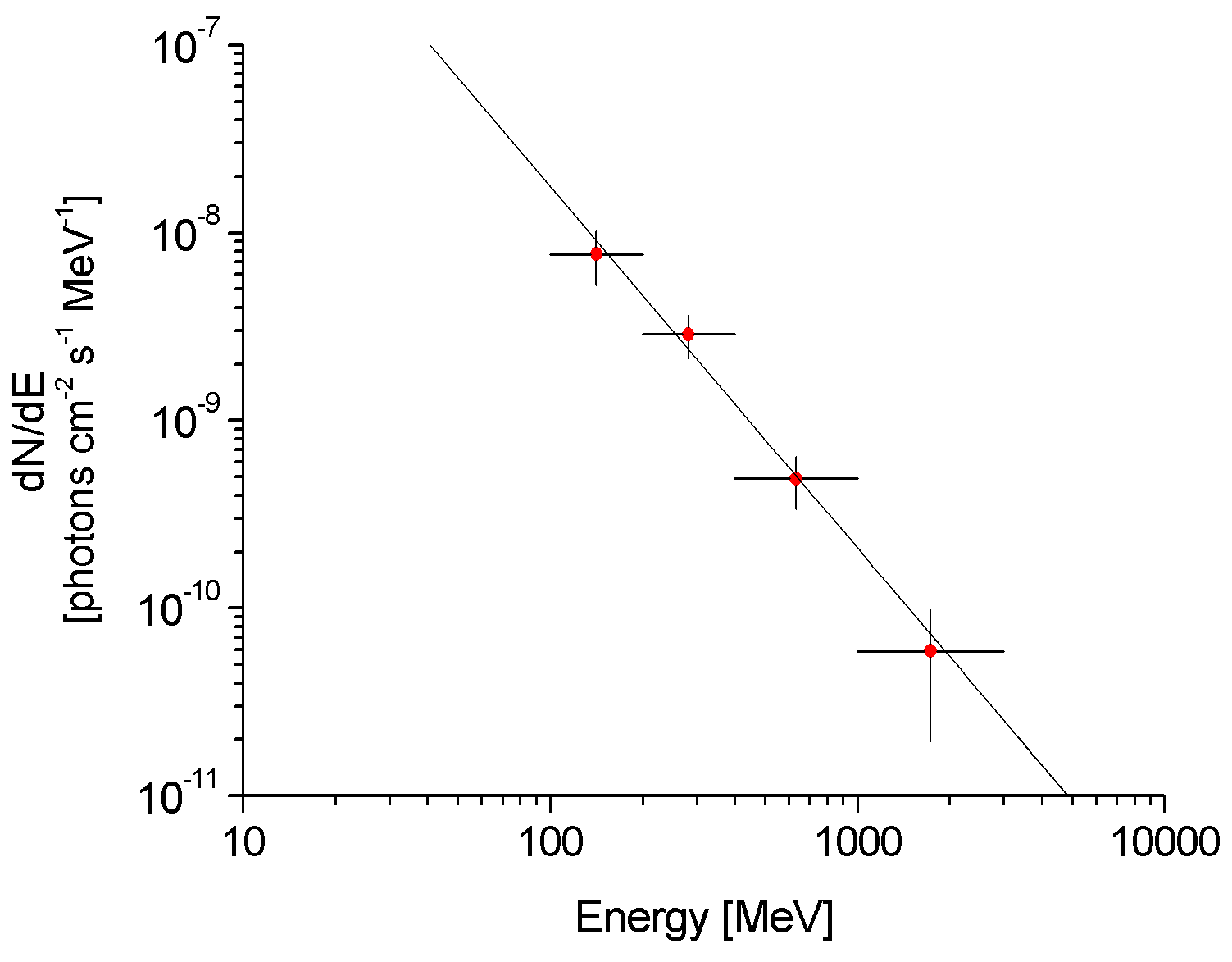} 
	\caption{\footnotesize Photon spectrum between 100 MeV and 3 GeV of Cygnus X-3 as detected by the \grid by integrating all flaring episodes in Table \ref{cyg_x3_all_flares}. Power law fit with photon index: $\Gamma=1.93$.} \label{cyg_x3_spectrum} 
\end{center}
\end{figure}

Between November 2007 and July 2009 AGILE repeatedly pointed at the Cygnus region for a total of $\sim$275 days, corresponding to a net exposure time of $\sim$11 Ms.
Seven \gray flares, each lasting 1-2 days, were detected (Table \ref{cyg_x3_all_flares}). The analysis was carried out by using a Multi-Source maximum-Likelihood Analysis (MSLA), to take into account the strong emission of the nearby \gray pulsars (PSR J2021+3651, PSR J2021+4026 and PSR J2032+4127).
By integrating all the flaring episodes, we found a \gray source detected at $5.9\sigma$ ($\sqrt{TS}=5.9$) at the average Galactic coordinate $(l,~b) = (80.0^{\circ},~0.9^{\circ})$ $\pm~0.5^{\circ}$ (stat) $\pm~0.1^{\circ}$ (syst), with a photon flux of [131 $\pm$ 27 (stat) $\pm$ 10\% (syst)]$\cdot 10^{-8}$ \flx. The average differential spectrum between 100 MeV and 3 GeV is well described by a power law with a photon index $\Gamma=1.93~\pm~0.23$ (Figure \ref{cyg_x3_spectrum}).
By using this photon index, a Multi-Source maximum-Likelihood Analysis (MSLA), applied in the deep integration of the \grid data (between November 2007 and July 2009), found a weak persistent emission (significance: $\sqrt{TS}=5.17$) from a position consistent with Cygnus X-3 with a photon flux of [14 $\pm$ 3  (stat) $\pm$ 10\% (syst)]$\cdot 10^{-8}$ \flx. So the average ``flaring'' flux is about 10 times the steady flux associated to Cygnus X-3, and the photon flux of a single \gray flare can be as high as $\sim$20 times greater than the steady flux (Table \ref{cyg_x3_all_flares}).

\begin{table}[h!]
\caption{\footnotesize Major \gray flares detected by the \grid in the period November 2007 - July 2009. All detections have a significance above $3\sigma$ ($\sqrt{TS} \geqslant 3$). \textit{Column one}: period of detection in MJD; \textit{Column two}: significance of detection; \textit{Column three}: photon flux.} \label{cyg_x3_all_flares}

\renewcommand{\arraystretch}{1.5}
{\footnotesize
\begin{tabular}{|c|r|c|}
\hline
MJD                  &  $\sqrt{\mathrm{TS}}$  &  Flux [$10^{-8}$ photons $\mathrm{cm^{-2}}$ $\mathrm{s^{-1}}$]\\
\hline\hline
 54507.76 - 54508.46 &           3.66         &           264  $\pm$  104  (stat)  $\pm$ 10\%  (syst)   \\

 54572.58 - 54573.58 &           4.48         &           265  $\pm$   80  (stat)  $\pm$ 10\%  (syst)   \\

 54772.54 - 54773.79 &           3.92         &           214  $\pm$   73  (stat)  $\pm$ 10\%  (syst)   \\

 54811.83 - 54812.96 &           3.98         &           190  $\pm$   65  (stat)  $\pm$ 10\%  (syst)   \\

 55002.88 - 55003.87 &           3.84         &           193  $\pm$   67  (stat)  $\pm$ 10\%  (syst)   \\

 55025.05 - 55026.04 &           3.23         &           216  $\pm$   89  (stat)  $\pm$ 10\%  (syst)   \\

 55033.88 - 55035.88 &           3.62         &           158  $\pm$   59  (stat)  $\pm$ 10\%  (syst)   \\
\hline
\end{tabular}
}
\end{table}

\section{Discussion}
\begin{figure*}
\begin{center}
	\includegraphics[width=18.0cm]{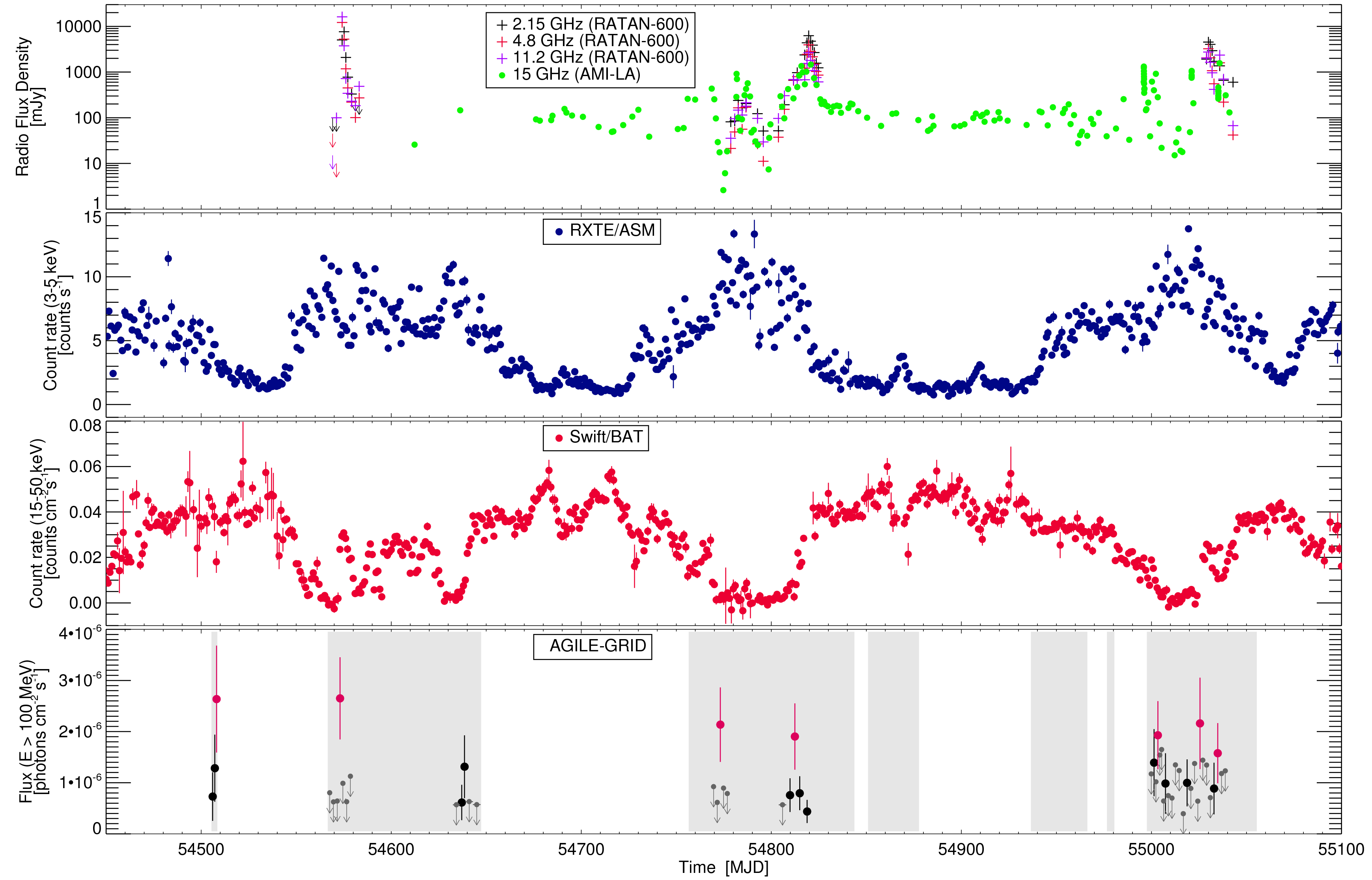} 
	\caption{\footnotesize Multi-frequency light curve of Cygnus X-3 from 2007-December-12 to 2009-September-26 (MJD: 54450-55100). From top to bottom: \textbf{radio} flux density [RATAN-600 (2.15, 4.8, 11.2 GHz) and AMI-LA (15 GHz)], \textbf{soft X-ray} count rate [\textit{RXTE}/ASM (3-5 keV)], \textbf{hard X-ray} count rate [\textit{Swift}/BAT (15-50 keV)] and $\boldsymbol\gamma$\textbf{-ray} photon fluxes [\grid (above 100 MeV)]. In the bottom panel gray regions represent the AGILE pointing at the Cygnus region; \textit{magenta} points are the \gray flares with $\sqrt{TS}\geqslant3$ (major \gray flares, see Table \ref{cyg_x3_all_flares}), \textit{black} points are the \gray detections with $2 \leqslant \sqrt{TS} < 3$ and \textit{dark-gray} arrows are $2\sigma$ the upper limits related to $\sqrt{TS} < 2$.} \label{cyg_x3_all_flares_mw} 
\end{center}
\end{figure*}

In order to analyze the pattern of emission, the plot in Figure \ref{cyg_x3_all_flares_mw} shows the comprehensive multi-wavelength light curve of Cygnus X-3. The \gray emission of the flaring episodes is compared with hard X-ray fluxes from \textit{Swift}/BAT (15-50 keV), soft X-ray fluxes from \textit{RXTE}/ASM (3-5 keV) and radio flux density (when available) from AMI-LA (15 GHz) and RATAN-600 (2.15, 4.8, 11.2 GHz) radio telescopes.

Observing the light curve in Figure \ref{cyg_x3_all_flares_mw}, we can notice that:

\begin{itemize}
  \item there is a \textit{strong anticorrelation} between hard X-ray and \gray emission: every local minimum of the hard X-ray light curve is associated with \gray emission detected by the \grid (see also the weak \gray event detected on 2008-June-21, MJD = 54638.58, in the plot of Figure \ref{cyg_x3_all_flares_mw}: $\sqrt{TS}=2.77$, photon flux = [131 $\pm$ 61 (stat) $\pm$ 10\% (syst)]$\cdot 10^{-8}$ \flx);
  vice versa, every time the \grid detects \gray activity -- with exception of the \gray flare detected on 11-12 February 2008  (MJD: 54507.76--54508.46) -- the system exhibits a very deep local minimum ($\leqslant 0.01~\mathrm{counts}$ $\mathrm{cm^{-2}~s^{-1}})$) of the hard X-ray light curve;
  \item every time we detect \gray activity, Cygnus X-3 is in a soft spectral state (the 3-5 keV \textit{RXTE}/ASM count rate $\gtrsim~3~\mathrm{counts~s^{-1}}$);
  \item every time we detect \gray flaring episodes (Table \ref{cyg_x3_all_flares} and red points in the \grid light curve in Figure \ref{cyg_x3_all_flares_mw}) -- with exception of the \gray flare of 11-12 February 2008 (MJD: 54507.76--54508.46) -- the system is moving to a major radio flare (radio flux density $\gtrsim1$ Jy) or to a quenched state preluding a major radio flare.
\end{itemize}

\begin{figure}[h!]
 \begin{center}
	\includegraphics[width=7.0cm]{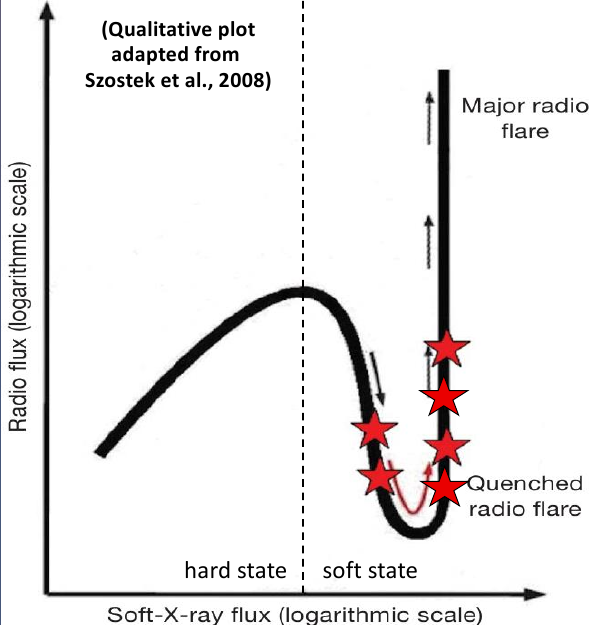} 
	\caption{\footnotesize Schematic representation of the evolution of Cygnus X-3 through its radio and X-ray state. The red stars mark the approximate position of the \gray flares detected by the \grid, occurring in pre-quenching/pre-flaring radio states.} \label{cyg_x-3_sax} 
\end{center}
\end{figure}

In summary, the \grid detected enhanced \gray emission when the system is in a bright soft X-ray spectral state, corresponding to a minimum of the hard X-ray emission, few days before a major radio flare (the “gully” of diagram in Figure \ref{cyg_x-3_sax}).

A separate discussion is needed for the \gray flare of 11-12 February 2008 (MJD: 54507.76--54508.46), a special event among the \grid detections. The AGILE satellite pointed at the direction of the blazar Markarian 421 (Mkn 421) for a Target of Opportunity (ToO) during the period 2008 February 09-12 (ToO Mkn 421, Observation Block 5210, Galactic coordinates of the pointing centroid: $(l,~b)~=~(77.3,~40.6)$). During this short period the \grid detected \gray activity from Cygnus X-3. Unfortunately we have no available radio dataset covering this period, so we do not have all the ingredients we need to deeply analyze the behavior of the microquasar. Furthermore, even if the soft X-ray flux is $\sim$3 $\mathrm{counts~s^{-1}}$ (i.e., Cygnus X-3 is on the transition level), the system does not appear to be in a bright soft spectral state: the average hard X-ray flux is quite high ($\sim$0.03-0.04 $\mathrm{counts~cm^{-2}~s^{-1}}$). Anyway, the \gray flare is coincident with a little but sharp dip ($\sim$0.01-0.02 $\mathrm{counts~cm^{-2}~s^{-1}}$) of the \textit{Swift}/BAT light curve. It seems to confirm the simultaneous \gray-flare/hard-X-ray-minimum occurrence that we find in all other cases.

\subsection*{Perspectives on spectral modeling}

A possible leptonic interpretation \citep{dubus_10} of the \gray emission is based on inverse Compton scattering of soft photons by high energy electrons of the jet. The HE \gray modulated emission could explained in a natural way by using a defined geometry of the jet model: a jet launched around the compact object with moderate bulk relativistic speed, oriented not too far from the line-of-sight, interacting with the WR star wind to produce a shock very close to the compact object; in this shock, electrons - accelerated to GeV energies - upscatter (via IC processes) soft photons to energies above 100 MeV. According to this phenomenological picture, Cygnus X-3 is a microblazar, with a jet pointing towards the Earth.

The strong \gray emission and the differential photon spectrum (Figure \ref{cyg_x3_spectrum}) detected by the the \grid during the flaring activity of Cygnus X-3 can be interpreted in a natural way by assuming this simple leptonic scenario based on IC scatterings by relativistic electrons (and positrons) injected in the jet structure. Our attempt to model the multiwavelength SED of Cygnus X-3, during the \gray flaring activity, will be published soon in a dedicated paper \citep{piano_11}. Our model takes into account the X-ray, \gray (\grid) and TeV emission from the microquasar during the peculiar spectral states when AGILE detected the high energy flares. This model requires the introduction of a new component (``IC bump'') in the SED of the system. The theoretical expectations based on hybrid-Comptonization models in the corona, commonly used to model the X-ray spectra of the microquasars, fail to properly explain the observed HE \gray flares.

\section{Conclusions}
Several major \gray flares were detected by the \grid from Cygnus X-3 while the system was in peculiar radio/X-ray spectral states: intense \gray activity was detected always during prominent minima of the hard X-ray flux (corresponding to strong soft X-ray emission), a few days before intense radio outbursts (major radio flares). This temporal repetitive coincidence turns out to be the \textit{spectral signature} of the \gray activity from this puzzling microquasar, and might open new areas in which to study the interplay between the accretion disk, the corona and the formation of relativistic jets.

\bigskip 

\end{document}